# Quantum Interference in a Single Perovskite Nanocrystal

Yan Lv,[1] Chunyang Yin,[1] Chunfeng Zhang,[1] William W. Yu,[2] Xiaoyong Wang,[1*] Yu Zhang,[2*] and Min Xiao[1,3*]

[1]*National Laboratory of Solid State Microstructures, School of Physics, and Collaborative Innovation Center of Advanced Microstructures, Nanjing University, Nanjing 210093, China*

[2]*State Key Laboratory of Integrated Optoelectronics and College of Electronic Science and Engineering, Jilin University, Changchun 130012, China*

[3]*Department of Physics, University of Arkansas, Fayetteville, AR 72701, USA*

*Correspondence to X.W. (wxiaoyong@nju.edu.cn), Y.Z. (yuzhang@jlu.edu.cn) or M.X. (mxiao@uark.edu)

**Coherent manipulation of the exciton wave function in a single semiconductor colloidal nanocrystal (NC) has been actively pursued in the past decades without any success, mainly due to the bothersome existences of the spectral diffusion and the photoluminescence (PL) blinking effects. Such optical deficiencies can be naturally avoided in the newly-developed colloidal NCs of perovskite $CsPbI_3$, leading to the PL spectrum with a stable intensity at the single-particle level. Meanwhile, from the first-order photon correlation measurement, a PL linewidth smaller than 20 μeV is estimated for the emission state of the neutral excitons in a single $CsPbI_3$ NC. Moreover, a dephasing time of about 10 ps can be extracted from the quantum interference measurement on the absorption state of the charged excitons. This first demonstration**

**of a coherent optical feature will advance single colloidal NCs into the quantum information regime, opening up an alternative yet prospective research direction beyond their traditional applications such as in optoelectronic devices and bioimaging.**

When a quantum system is resonantly driven by two laser pulses with an increasing time delay, the wave functions generated with a correlated phase could interfere with each other to cause a population oscillation between the two-level states. This quantum interference effect, originally serving as a hallmark signature for single natural atoms[1], was later demonstrated in the absorption state of a single epitaxial quantum dot (QD) by monitoring the photoluminescence (PL) intensity variations of its emission state[2]. The subsequent explorations of various other coherent optical properties, such as Rabi oscillation[3], vacuum Rabi splitting[4] and Mollow triplet[5], have firmly pushed single epitaxial QDs to the fundamental and practical contexts of quantum information processing. In contrast, the coherent optical measurements of single colloidal nanocrystals (NCs) have never been directly implemented mainly due to their sufferings from the spectral diffusion[6,7] and the PL blinking[8] effects, which would reduce the dephasing time of the absorption state and fluctuate the PL intensity of the emission state, respectively. To minimize these optical instabilities, the emission-state spectrum of a single colloidal CdSe NC had to be acquired with an integration time shorter than 100 ms by means of the photon-correlation Fourier spectroscopy[9] or the resonant PL excitation technique[10], yielding a PL linewidth narrower than 10 μeV that could be indirectly converted to an exciton dephasing time of at least 100 ps. It is obviously challenging that a comparable dephasing time could be obtained from the quantum interference measurement normally lasting for tens of minutes, so that single colloidal NCs would truly step into the artificial-atom regime with previously unexplored coherent optical properties.

Compared to traditional metal-chalcogenide CdSe NCs that have been intensively studied for more than two decades, low-dimensional perovskites of caesium lead halide are just emerging as a novel type of colloidal NCs with superior optical properties. Here we perform PL and PL excitation measurements of single $CsPbI_3$ NCs at the cryogenic temperature of 3 K to reveal the emission and absorption states of their neutral and charged excitons, respectively. For the emission state of the neutral excitons, a PL linewidth less than 20 μeV can be estimated from the first-order photon correlation measurement of a single $CsPbI_3$ NC, which is almost twice narrower than that measured for the charged excitons. For the absorption state of the charged excitons, a dephasing time of about 10 ps is extracted from the quantum interference measurement, which signifies the first step towards exploring the coherent optical properties of single colloidal NCs.

The colloidal perovskite $CsPbI_3$ NCs with a cubic edge length of ~9.3 nm were chemically synthesized according to a standard procedure reported previously[11], with their ensemble PL spectrum showing a peak wavelength of 690 nm at the room temperature. One drop of the diluted NC solution was spin-coated onto a fused silica substrate, which was later attached to the cold finger of a helium-free cryostat operated at 3 K. The laser beam was focused by a dry objective (N.A. = 0.82) onto the sample substrate, and the PL signal collected by the same objective from a single $CsPbI_3$ NC could be sent through a spectrometer to either a charge-coupled-device (CCD) camera for the PL spectral measurement or an avalanche photodiode for the PL decay measurement with a time resolution of ~200 ps. The wave plate and linear polarizer would be added to the optical path for the laser excitation and/or PL collection whenever it was necessary for the polarization-dependent measurement. To avoid any nonlinear influence of multiple excitons in a single CsPbI$_3$ NC, we normally utilized a low laser power in our optical studies corresponding to an exciton number of at most 0.1 at a given time unless otherwise specified in the text. At this exciton number and with a 5.6 MHz laser excitation, the average photon count rate measured on an avalanche photodiode for a single CsPbI$_3$ NC was ~6 × 10$^2$ s$^{-1}$.

In Fig. 1a, we plot the PL spectrum of a single CsPbI$_3$ NC excited at 590 nm with a 5.6 MHz picosecond fiber laser, where the two peaks originate from the optical emission of the neutral excitons (*X*s) with orthogonal and linear polarizations[11]. These two peaks possess a good optical stability against the measurement time as can be seen from the PL spectral image plotted in Fig. 1b with only a slight occurrence of spectral diffusion. The appearance of such a PL doublet from the *X*s is still under active investigation and has been attributed to the electron-hole exchange interaction[11,12] or the Rashba effect[13,14] in a single perovskite NC. In Fig. 1c and d, we also plot respectively the PL spectrum and the time-dependent PL spectral image of the charged excitons (*X*$^-$s) for a single CsPbI$_3$ NC, which demonstrate a single peak due to the eliminated electron-hole exchange interaction[11]. PL transition from the *X* to the *X*$^-$ state could be induced by an increase of the laser excitation power, which was then lowered down to leave a single CsPbI$_3$ NC permanently dwelling at the charged state. While all of the optical measurements in Fig. 1 were performed at the exciton number of ~0.1, similar results are also shown in Supplementary Fig. S1 for single CsPbI$_3$ NCs excited at the exciton number of ~1.0.

The PL linewidths of the *X* and *X*$^-$ peaks shown in Fig. 1 are mainly limited by our spectral resolution of ~100 μeV, while a precise estimation can be realized by the first-order photon correlation measurement that have been widely applied to single epitaxial QDs[15,16]. In the following experiment, a single CsPbI$_3$ NC was excited at 632 nm by a continuous-wave He-Ne laser with the emitted photons being sent to a Michelson interferometer placed before the entrance of the spectrometer. One optical path of the interferometer was mechanically varied to generate a coarse time delay $\tau_c$ with a step resolution of 6.67 ps, while a fine time delay $\tau_f$ could be further imposed at each $\tau_c$ with a 0.2 fs step resolution. We first studied the $X$s of a single CsPbI$_3$ NC whose PL spectral images of the doublet peaks at two $\tau_c$ positions of 0 and 26.67 ps are plotted in Fig. 2a as a function of $\tau_f$ from 0-20 fs. These doublet PL peaks presented very similar interferometric correlation behaviours, so that we will only focus on the lower-energy one whose PL intensity was maximized through a polarized detection. In the inset of Fig. 2b where $\tau_c$ was set at 46.67 ps and $\tau_f$ was changed from 0-5 fs, a clear interference pattern could be observed for the PL intensity acquired in the CCD camera at each $\tau_f$. By fitting with a sinusoidal function, the maximum and minimum PL intensities of $I_{max}$ and $I_{min}$ could be extracted to calculate the visibility of ($I_{max} - I_{min}$)/($I_{max} + I_{min}$) at a given $\tau_c$[16]. In Fig. 2b, we plot this interference visibility as a function of $\tau_c$, which can be well fitted by a single-exponential dephasing time ($T_2$) of 75.86 ps.

The interferometric correlation measurements were also performed on the $X^-$s of a single CsPbI$_3$ NC, with the $\tau_f$ dependence of the PL spectral images being plotted in Fig. 2c at two $\tau_c$ positions of 0 and 13.34 ps, respectively. The corresponding intensity interference is shown in the inset of Fig. 2d at the $\tau_c$ position of 13.34 ps for the variation of $\tau_f$ from 0-5 fs. From the $\tau_c$ dependence of the interference visibility plotted in Fig. 2d, a $T_2$ value of 31.27 ps can be extracted from a single-exponential fitting. Similar $T_2$ values for the $X$s and $X^-$s were obtained for all the other single CsPbI$_3$ NCs studied in our experiment (see Supplementary Fig. S2). From the relationship of $\Gamma T_2 = h/\pi$, where $h$ is the Planck constant, the PL linewidth $\Gamma$ of the $X$s ($X^-$s) for the single CsPbI$_3$ NC studied in Fig. 2a and b [Fig. 2c and d] is estimated to be 17.4 (42.1) μeV. These PL linewidths are obviously broader than that of ~1.3 μeV calculated from the radiative lifetime $T_1$ of ~1 ns measured for the $X$s and $X^-$s (see Supplementary Fig. S3), implying that they are still suffering from a residual influence of the spectral diffusion effect. However, the PL linewidth of 17.4 μeV measured here for the $X$s of a single CsPbI$_3$ NC has represented a significant improvement of almost one order of magnitude over those best values ever reported for single CdSe NCs from the time-integrated PL measurements[6].

The existence of an ultranarrow linewidth in the emission state of a single CsPbI$_3$ NC should be also manifested in its absorption state, which can be probed by the PL excitation measurement. Previously, an absorption-state linewidth as broad as 3-4 meV was revealed from the PL excitation measurement performed on single colloidal CdSe NCs at the cryogenic temperature, where the PL blinking influence had to be intricately compensated by a special spectroscopic technique[17]. Due to the good optical stability of single CsPbI$_3$ NCs, it was easy for us to directly map their absorption states by employing a picosecond Ti:sapphire laser with a 76 MHz repetition rate. To avoid the scattered excitation light, the tunable laser was always set at an initial wavelength ~2 nm blue-shifted relative to that of the PL peak and changed towards even shorter wavelengths with a step resolution of ~0.02 nm. In Fig. 3a, we plot the PL and PL excitation spectra measured for the $X^-$s of a representative CsPbI$_3$ NC, where two absorption states with resolution-limited linewidths of ~200 μeV are respectively resolved at higher energies of ~16.9 and ~19.8 meV than that of the emission state. For the $X^-$ emission state of another single CsPbI$_3$ NC shown in Fig. 3b, only one absorption state shows up at a higher energy of ~11.6 meV. Moreover, a continuous distribution of absorption states could be detected for the $X^-$s of a single CsPbI$_3$ NC when the excitation laser was tuned even farther away from the emission state (see Supplementary Fig. S4a). Compared to the $X^-$s, the PL excitation spectrum measured for the $X$s is more complicated, displaying both discrete and clustered absorption states (see Supplementary Fig. S4b). Although the exact origins of these absorption states are yet to be determined in future works, their discrete existence especially in the $X^-$s makes it possible to attempt the quantum interference measurement on a single CsPbI$_3$ NC.

In the next experiment, we tuned the output wavelength of the above Ti:sapphire laser to a discrete $X^-$ absorption line of a single CsPbI$_3$ NC and sent the beam through a Michelson interferometer to generate two pulse trains with coarse ($\tau_c$) and fine ($\tau_f$) time delays of 2 ps and 0.2 fs step resolutions, respectively. In Fig. 4a, we first plot the autocorrelation function measured for the excitation laser whose $I_{max}$ and $I_{min}$ change exponentially with the increasing $\tau_c$ to yield a fitted pulse width of 3.95 ps. The $I_{max}$ and $I_{min}$ measured for the $X^-$s at each $\tau_c$ for a single CsPbI$_3$ NC are also presented in Fig. 4a and each fitted by a single-exponential function of $\tau_c$ to yield a dephasing time $T_2$ of 9.25 ps. In the inset of Fig. 4b, we plot the $\tau_f$ dependence of the PL intensity oscillation measured for a single CsPbI$_3$ NC, which serves as an example to show how $I_{max}$ and $I_{min}$ were obtained at a specific $\tau_c$ position of 12 ps. This sinusoidal oscillation of the PL intensity, at a time delay when the temporal overlap of the two laser pulse trains is absent, unambiguously signifies the quantum interference of the exciton wave functions in the $X^-$ absorption state. The interference visibility of $(I_{max} - I_{min})/(I_{max} + I_{min})$ calculated at each $\tau_c$ position for this single CsPbI$_3$ NC is shown in Fig. 4b, while a single-exponential fitting leads to a $T_2$ value of 11.12 ps. For about ten single CsPbI$_3$ NCs studied in our experiment, an average $T_2$ value of 9.83 ps could be obtained with the longest and shortest $T_2$'s being 12.75 and 6.40 ps, respectively (see Supplementary Fig. S5 for two more single CsPbI$_3$ NCs). The $T_2$ values for the absorption state of $X^-$s are obviously shorter than those obtained from the first-order photon correlation measurement of the emission state, which can be naturally explained by a further reduction of the coherence time due to the exciton relaxation process in addition to the spectral diffusion effect.

To summarize, we have performed several optical measurements on single perovskite CsPbI$_3$ NCs that are otherwise inaccessible for traditional colloidal NCs. Besides the first-order photon correlation and the PL excitation measurements, the quantum interference effect is successfully demonstrated for the $X^-$ absorption state of a single CsPbI$_3$ NC with a dephasing time of ~10 ps, which is on par with those values obtained from the earlier studies of this coherent optical phenomenon in single epitaxial QDs[18-20]. The above measurements on the CsPbI$_3$ NCs rely critically on their superior optical stabilities in the PL intensity and linewidth, which have been similarly demonstrated in other types of colloidal perovskite NCs such as CsPbBr$_3$[12,13,21,22] and CsPbBr$_2$Cl[14] as signified by their exciton fine-structure splittings. Thus, this quick addition of the quantum interference feature to the already-intriguing optical properties will surely stimulate intensive coherent optical investigations on the newly-emerged colloidal perovskite NCs with various structures and compositions[23]. For the current colloidal NC research, the synthesis procedures are being continuously polished to almost exclusively suit their practical applications in the optoelectronic devices, while the first revelation of a coherent optical property here has provided another potent outlet towards their ultimate usages in coherent computing and quantum information processing. In terms of the cheap and flexible synthesis, the broad tunability across the whole visible spectrum, as well as the easy integration into different optical structures, the colloidal perovskite NCs are possessing obvious advantages over the traditional epitaxial QDs. It can be naturally envisioned that the colloidal perovskite NCs are capable of serving as an alternative platform to attract researchers from the epitaxial-QD community, so that their well-developed characterization tools would be applied to reproduce the old and to explore the new coherent optical properties of semiconductor nanostructures.

For the following coherent optical studies of the single perovskite NCs, we are optimistic that a longer dephasing time could be achieved from the emission state by means of its resonant excitation in a microcavity to suppress the scattered laser light[24]. It has been theoretically proposed that the energy level of dark excitons in a single perovskite NC is above those of the bright excitons[14], so their efficient generation, possibly through a two-photon excitation scheme, and intrinsic slow recombination[25] would provide another opportunity for the acquirement of an elongated dephasing time. Moreover, the easy access to the $X^-$s with an extra charge in a single perovskite NC and the ability to dope it with impurity ions[26] could also be utilized to render a long-term storage of the coherence information in the spin state[27].

**Acknowledgements**

This work is supported by the National Key R&D Program of China grant 2017YFA0303700, the National Natural Science Foundation of China grants 11574147 and 11621091, and the PAPD of Jiangsu Higher Education Institutions.

**Author contributions:**

X.W. and M.X. proposed and supervised the project; Y.L. and C.Y. performed the optical measurements; W.Y. and Y.Z. synthesized the samples; Y.L. and X.W. analyzed the data; X.W. C.Z. and M.X. co-wrote the manuscript; and all authors discussed the results and commented on the manuscript.

**Additional information**

The authors declare no competing financial interests. Correspondence and requests for materials should be addressed to X.W., Y.Z. or M.X.

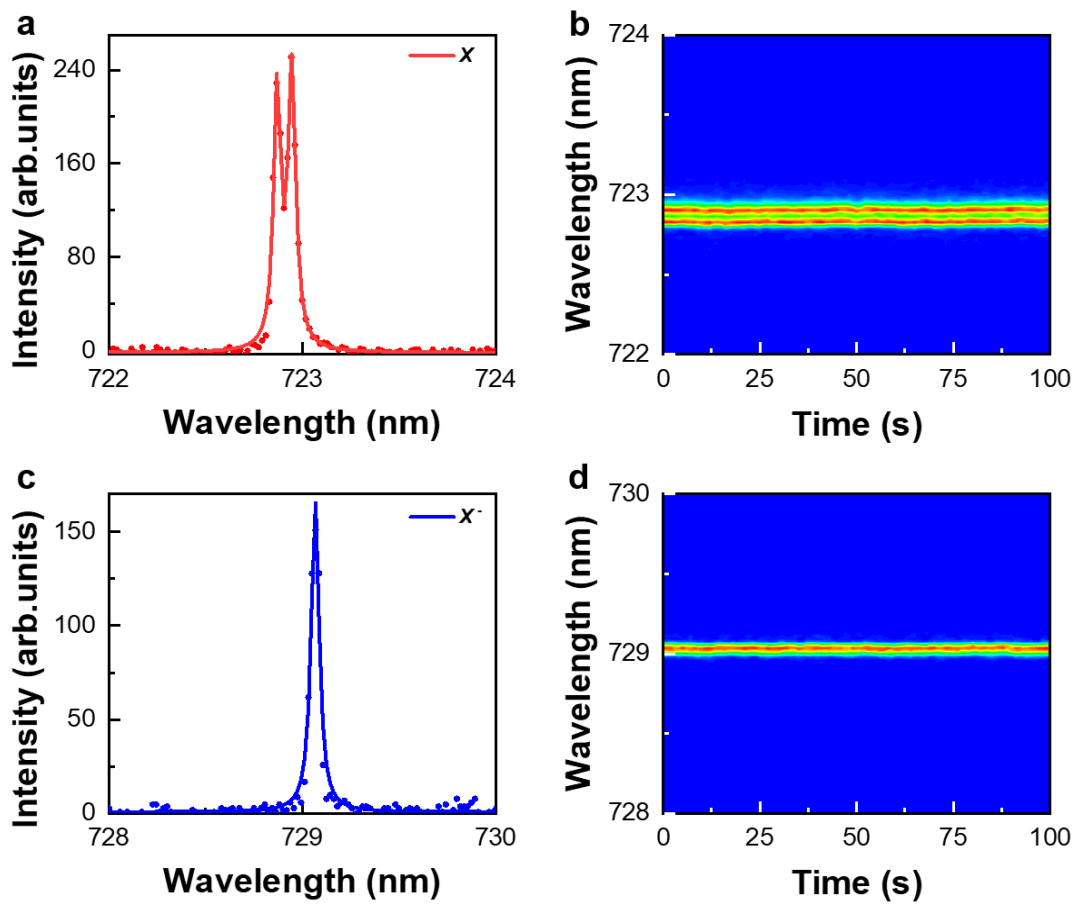

**Figure 1 | Emission states of single CsPbI$_3$ NCs. a,** PL spectrum and **b,** time-dependent PL spectral image of the *X*s with doublet peaks in a single CsPbI$_3$ NC. **C,** PL spectrum and **d,** time-dependent spectral image of the *X*⁻s with a single peak in a single CsPbI$_3$ NC. All these PL spectra were acquired with an integration time of 1 s with unpolarized excitation and detection. The single CsPbI$_3$ NCs were excited at 590 nm with a 5.6 MHz picosecond fiber laser, and their optical signals were separated from the excitation laser by a long-pass optical filter.
13

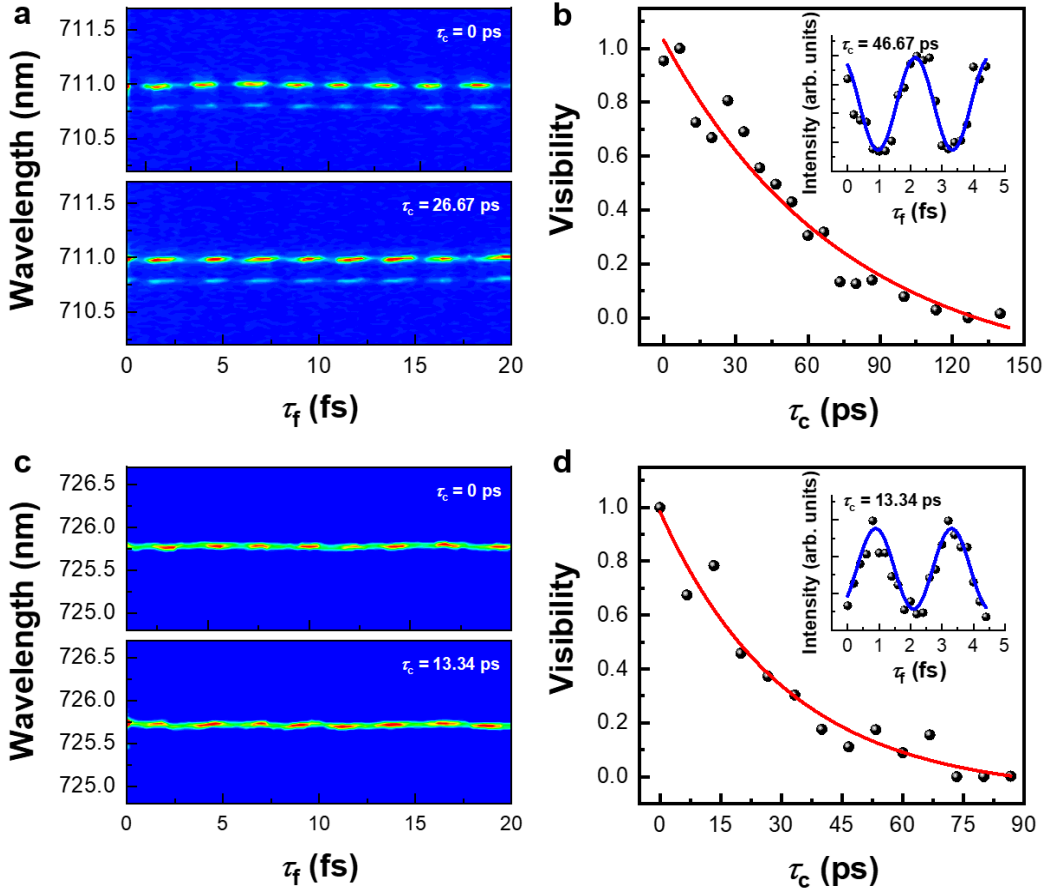

**Figure 2 | First-order photon correlation measurements of single CsPbI$_3$ NCs. a,** 2D plot in the wavelength-time plane for the PL intensity oscillations of the *X*s in a single CsPbI$_3$ NC as a function of $\tau_f$ at $\tau_c$ = 0 and 26.67 ps, respectively. The PL intensity for one of the doublet peaks was maximized through the polarized detection. **b,** $\tau_c$ dependence of the *X* interference visibility fitted with a $T_2$ value of 75.86 ps. Inset: Expanded view at $\tau_c$ = 46.67 ps showing the $\tau_f$ dependence of the *X* PL intensity oscillation. **c,** 2D plot in the wavelength-time plane for the PL intensity oscillations of the *X*⁻s in a single CsPbI$_3$ NC as a function of $\tau_f$ at two $\tau_c$ positions of 0 and 13.34 ps, respectively. **d,** $\tau_c$ dependence of the *X*⁻ interference visibility fitted with a $T_2$ value of 31.27 ps. Inset: Expanded view at $\tau_c$ = 13.34 ps showing the $\tau_f$ dependence of the *X*⁻ PL intensity oscillation. The PL spectrum measured at each $\tau_f$ for the construction of the 2D plots in **a** and **c** was acquired with an integration time of 1 s. The single CsPbI$_3$ NCs were excited at 632 nm with a continuous-wave He-Ne laser, and their optical signals were separated from the excitation laser by a long-pass optical filter.

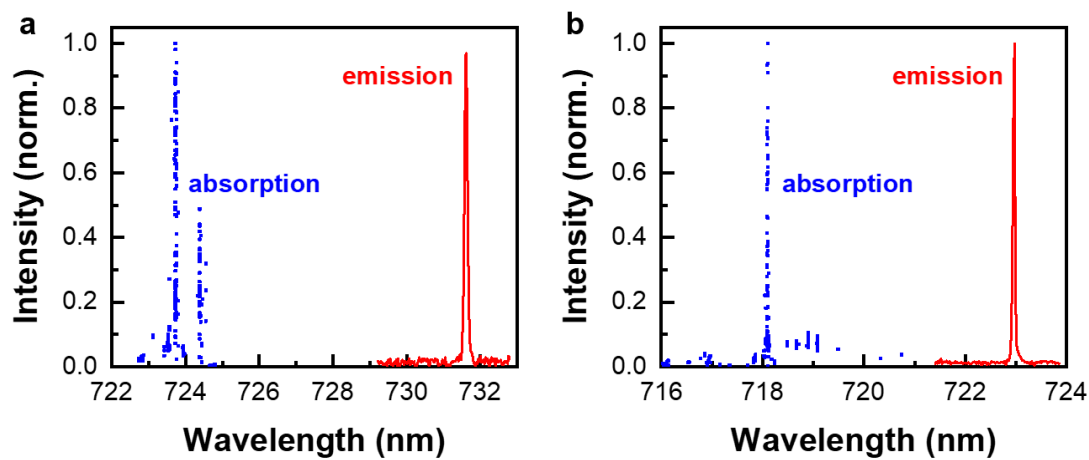

**Figure 3 | PL excitation measurements of single CsPbI$_3$ NCs. a,** PL and PL excitation spectra measured for the $X^-$s of a single CsPbI$_3$ NC with two absorption states. **b,** PL and PL excitation spectra measured for the $X^-$s of a single CsPbI$_3$ NC with one absorption state. The data points for constructing the PL excitation spectrum were obtained from several scans of the tunable laser wavelength in order to precisely determine the peak position of the absorption state. An integration time of 1 s was employed for acquiring the PL spectrum of the emission state in the PL excitation measurement. When the single CsPbI3 NCs were excited by the Ti:sapphire laser, their optical signals were separated from the excitation laser by two tunable long-pass optical filters and the 1200 g/mm grating equipped on a 0.75 m long spectrometer.

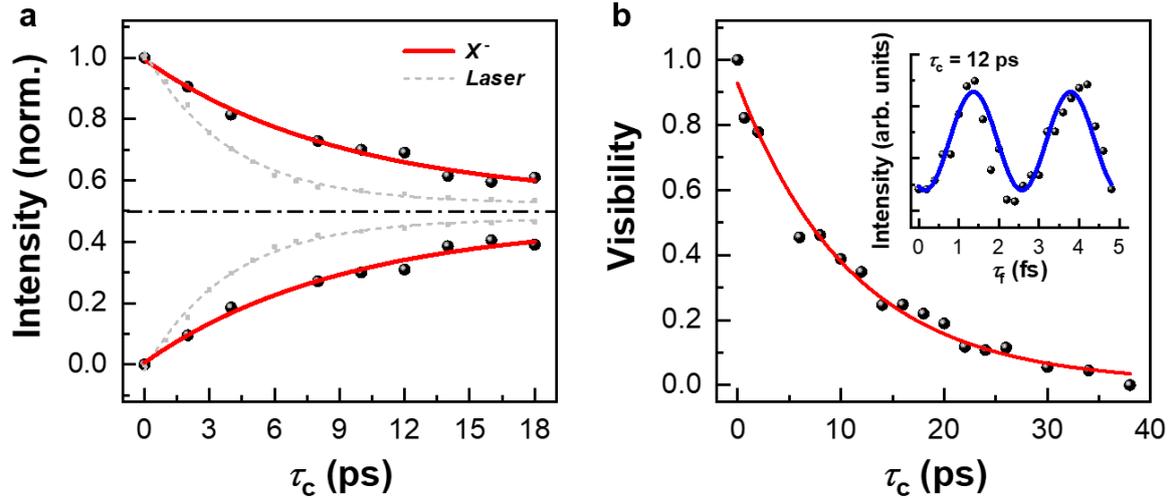

**Figure 4 | Quantum interference measurements of single CsPbI$_3$ NCs. a,** Maximum (top) and minimum (bottom) PL intensities measured for the $X^-$s of a single CsPbI$_3$ NC as a function of $\tau_c$ and each fitted with a single-exponential function to yield a $T_2$ value of 9.25 ps. The grey dotted lines correspond to the autocorrelation function of the excitation pulses. The black dashed-dotted line marks the 0.5 position for the normalized PL intensity. **b,** $\tau_c$ dependence of the $X^-$ interference visibility measured for a single CsPbI$_3$ NC and fitted with a single-exponential function to yield a $T_2$ value of 11.12 ps. Inset: Expanded view at $\tau_c = 12$ ps showing the $\tau_f$ dependence of the $X^-$ PL intensity oscillation. The PL spectrum was measured at each $\tau_f$ with an integration time of 1 s.

# SUPPLEMENTARY INFORMATION

**Quantum Interference in a Single Perovskite Nanocrystal**

Yan Lv,[1] Chunyang Yin,[1] Chunfeng Zhang,[1] William W. Yu,[2] Xiaoyong Wang,[1]* Yu Zhang,[2]* and Min Xiao[1,3]*

[1]*National Laboratory of Solid State Microstructures, School of Physics, and Collaborative Innovation Center of Advanced Microstructures, Nanjing University, Nanjing 210093, China*

[2]*State Key Laboratory of Integrated Optoelectronics and College of Electronic Science and Engineering, Jilin University, Changchun 130012, China*

[3]*Department of Physics, University of Arkansas, Fayetteville, AR 72701, USA*

*Correspondence to X.W. (wxiaoyong@nju.edu.cn), Y.Z. (yuzhang@jlu.edu.cn) or M.X. (mxiao@uark.edu)

.

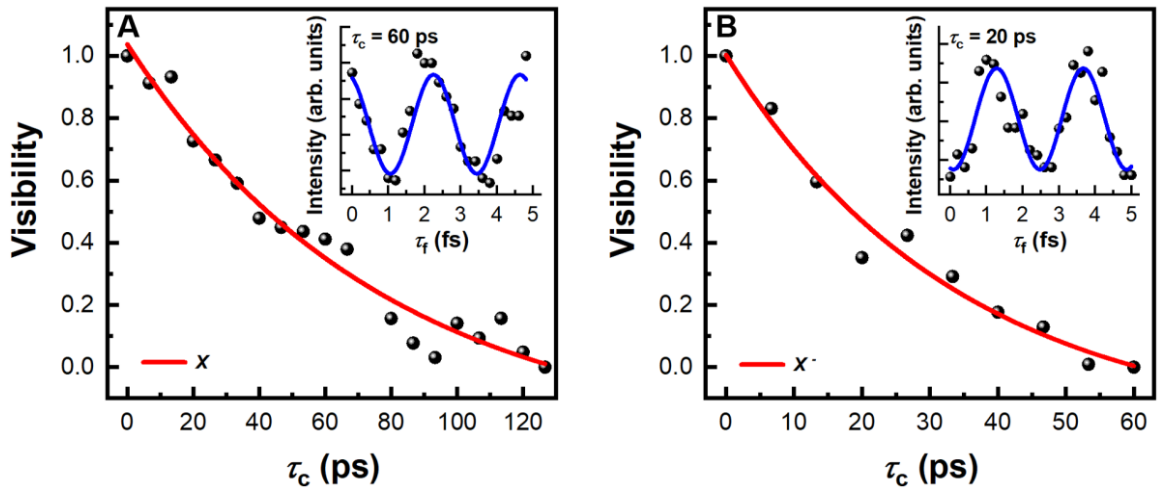

**Fig. S1. First-order photon correlation measurements of single CsPbI$_3$ NCs.** (A) $\tau_c$ dependence of the *X* interference visibility measured for a single CsPbI$_3$ NC and fitted with a $T_2$ value of 77.83 ps. Inset: $\tau_f$ dependence of the *X* PL intensity oscillation measured at $\tau_c$ = 60 ps. (B) $\tau_c$ dependence of the *X⁻* interference visibility measured for a single CsPbI$_3$ NC and fitted with a $T_2$ value of 34.45 ps. Inset: $\tau_f$ dependence of the *X⁻* PL intensity oscillation measured at $\tau_c$ = 20 ps.

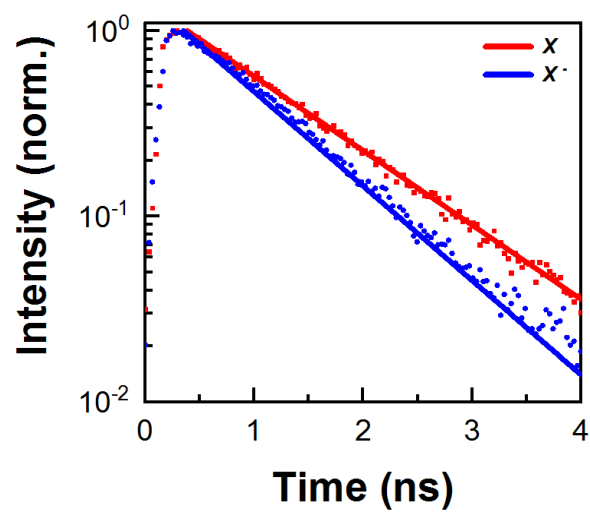

**Fig. S2. PL decay curves of a single CsPbI$_3$ NC.** The $X$ and $X^-$ PL lifetimes were obtained from single-exponential fittings to be 1.09 and 0.85 ns, respectively.

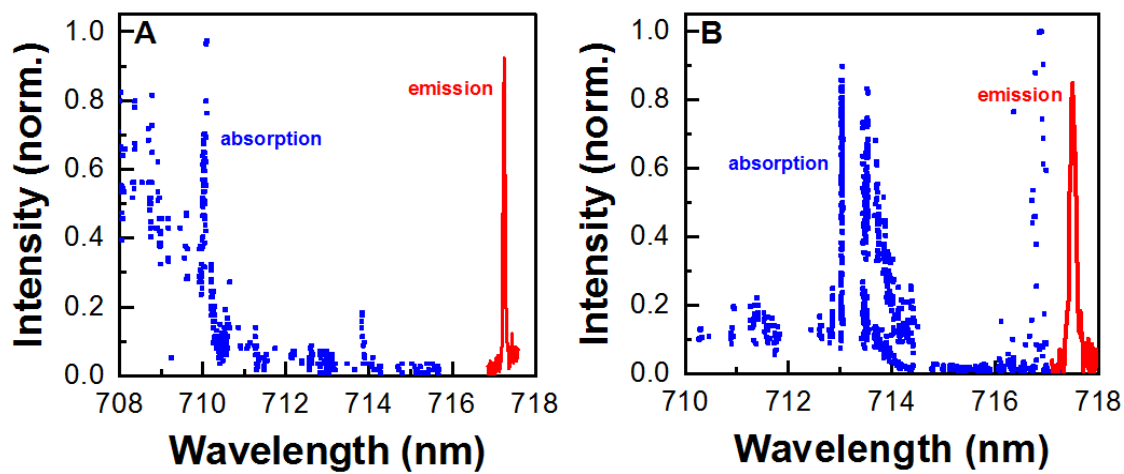

**Fig. S3. PL excitation measurements of single CsPbI$_3$ NCs.** (**A**) PL and PL excitation spectra measured for the $X^-$s of a single CsPbI$_3$ NC. (**B**) PL and PL excitation spectra measured for the *X*s of a single CsPbI$_3$ NC with polarized excitation and detection of only one of the doublet peaks. The data points for constructing each PL excitation spectrum were obtained from several scans of the tunable laser wavelength.

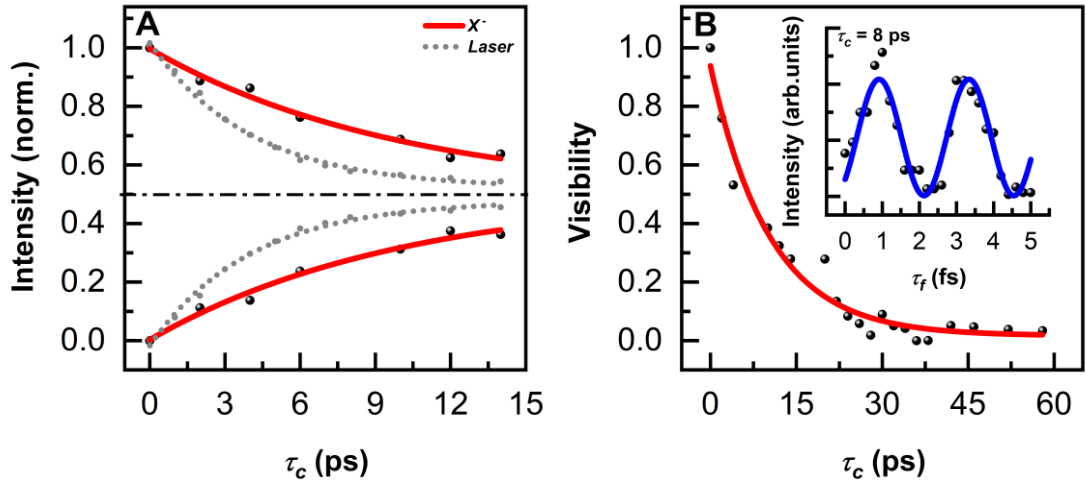

**Fig. S4. Quantum interference measurements of single CsPbI$_3$ NCs.** **(A)** Maximum (top) and minimum (bottom) PL intensities measured for the $X^-$s of a single CsPbI$_3$ NC as a function of $\tau_c$ and each fitted with a single-exponential function to yield a $T_2$ value of 10.13 ps. The grey dotted lines correspond to the autocorrelation function of the excitation pulses. The black dashed-dotted line marks the 0.5 position of the normalized PL intensity. **(B)** $\tau_c$ dependence of the $X^-$ interference visibility measured for a single CsPbI$_3$ NC and fitted with a single-exponential function to yield a $T_2$ value of 10.36 ps. Inset: $\tau_f$ dependence of the $X^-$ PL intensity oscillation measured at $\tau_c = 8$ ps.